\documentclass{article}

\usepackage[left=3cm, right=3cm, top=2cm]{geometry} 
\usepackage[textsize={footnotesize } 
]{todonotes}

\begin{document}
\title{Anisoplanatic adaptive optics in parallelized laser scanning microscopy}
\date{}

\author{Paolo Pozzi,$^{1,*}$ Carlas Smith,$^1$ Dean Wilding,$^1$  Oleg Soloviev,$^{1,2}$, Martin Booth$^3$,\\  Gleb Vdovin$^{1,2}$ and Michel Verhaegen$^1$}

\maketitle

\noindent $^{1}$ Delft Center for Systems and Control, Delft University of Technology, Mekelweg 2, 2628 CD Delft, The Netherlands\\
$^{2}$ Flexible Optical B.V., Polakweg 10-11, 2288 GG Rijswijk, The Netherland\\
$^{3}$ Department of Engineering Science, University of Oxford, Parks Road, Oxford OX1 3PJ, UK

e-mail:*p.pozzi@tudelft.nl

\begin{abstract}
Inhomogeneities in the refractive index of a biological sample can introduce phase aberrations in microscopy systems, severely impairing the quality of images.

Adaptive optics can be employed to correct for phase aberrations and improve image quality. However, conventional adaptive optics can only correct a single phase aberration for the whole field of view (isoplanatic correction) while, due to the three dimensional nature of biological tissues, the sample induced aberrations in microscopy often vary throughout the field of view (anisoplanatic aberration), limiting significantly the effectiveness of adaptive optics.

This paper reports on a new approach for aberration correction in laser scanning confocal microscopy, in which a spatial light modulator is used to generate multiple excitation points in the sample to simultaneously scan different portions of the field of view with completely independent correction, achieving anisoplanatic compensation of sample induced aberrations.

The method was tested in $150 \mu m$ thick whole Drosophila brains, showing a dramatic improvement in resolution and sharpness when compared to conventional isoplanatic adaptive optics. 
\end{abstract}

\section{Introduction}
Adaptive optics(AO) is a method of growing popularity for the correction of phase aberrations in optical systems.
Optical microscopes can particularly benefit from AO, as most samples of interest for life sciences are thick and inhomogeneous, introducing significant phase aberrations to light propagating through them. Implementations of AO have been reported for many of the most popular fluorescence microscopy techniques in life sciences \cite{ji2017adaptive}.

In the classical implementation of AO \cite{booth2002adaptive}, a two-dimensional phase modulator (2D-PM), such as a deformable mirror, or a liquid crystal spatial light modulator, is generally introduced in the back aperture plane of the system. In order to improve the image quality, the difference between the phase modulation introduced by the 2D-PM and the phase delay accumulated by light traveling through the sample should be minimized.

The main limitation in this approach is due to the fact that a 2D-PM in the pupil plane of an optical system is only capable of applying an isoplanatic correction, meaning the same two dimensional pupil plane correction is applied to all points within the field of view.

\begin{figure}[ht]
\centering
\includegraphics[width=\textwidth]{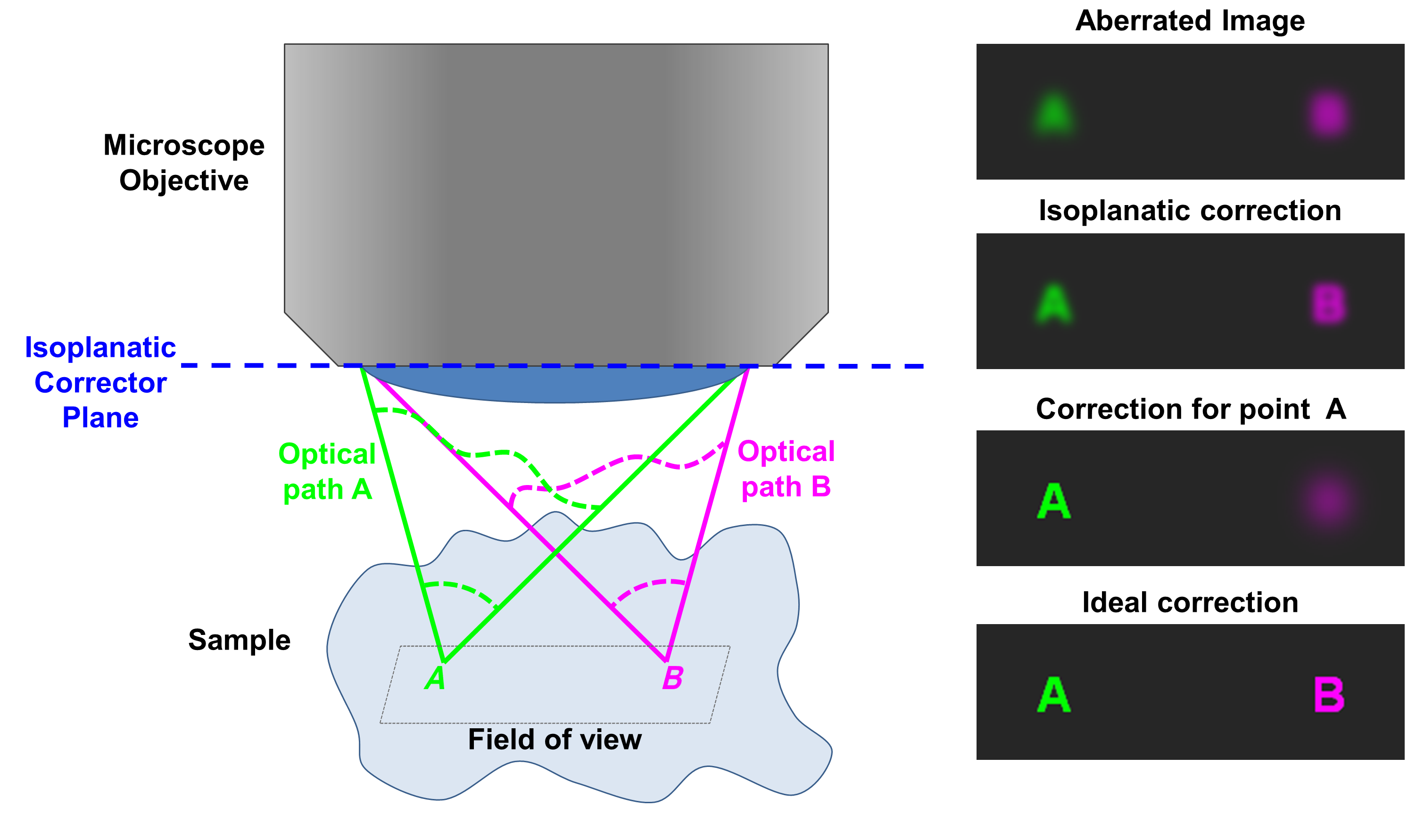}
\caption{\textbf{Anisoplanatic aberrations.} Schematic, not to scale representation of the problem of aberrations anisoplanatism in microscopy applications. Light from different parts of the field of view travel through different sections of the sample, and are therefore affected by different aberrations, which can not be compensated completely by a single corrector in the pupil plane.}
\label{fig: qualitativeanisoplanatism}

\end{figure}

The anisoplanatic nature of sample induced aberrations requires independent correction for each point of the field of view, which can not be achieved with a single 2D-PM, as schematically represented in figure \ref{fig: qualitativeanisoplanatism}.

A similar problem is encountered in astronomy, where multiple layers of turbulent atmosphere can affect the image quality, and the most common approach to correction is the use of multiple 2D-PM conjugated to different optical planes \cite{johnston1994analysis,marchetti2003mad}.

However, to the knowledge of the authors, this approach has never been reported in microscopy applications, possibly due to the high cost, technical difficulty of the implementation and the substantial control challenges involved.

A substantial improvement in anisoplanatic correction performances can be obtained through conjugate AO \cite{mertz2015field,park2015high}, in which the 2D-PM conjugation plane can be dynamically positioned in any plane between the microscope objective plane and the focal plane. However, while greatly improving the effectiveness of aberration correction, this approach only provides complete correction if the aberration is mostly confined to a 2D plane. Moreover, the variable magnification of such a system imposes a tradeoff between the size of the independently corrected patches and the order of the aberration. Similar constraints apply to multiplexed pupil methods \cite{park2017large}, in which different subregions of the 2D-PM are used to compensate aberrations in different areas of the field of view.
As a last possibility, anisoplanatic correction can be performed by imaging multiple fields of view of small dimension, independently applying correction for each of them, and combining the acquired images in a full field image \cite{liu2018observing}. This approach however can be extremely time consuming, especially in the case of laser scanning systems, while performing sensorless correction.

This paper presents a novel approach to anisoplanatic correction in laser scanning confocal fluorescence microscopy, allowing for spatially varying aberration correction of the excitation light throughout the field of view through the use of a liquid crystal spatial light modulator (SLM) and a low order deformable mirror (DM), both conjugated with the pupil plane of the system. The method allows simultaneous and independent aberration correction in up to hundreds of subregions of the field of view, without any tradeoff between number of patches and the resolution of the applied phase correction.

\section{Method description}

This paper introduces a new imaging method, named Laser Scanning Fluorescence Microscopy with Anisoplanatic Correction of Excitation (LSFM-ACE).
The confocal imaging principle of an LSFM-ACE microscope is equivalent to that of a conventional laser scanning confocal microscope, where excitation light is focused in the sample in a diffraction limited spot, and the confocal fluorescence emission is filtered through a pinhole.

Unlike a regular laser scanning microscope, LSFM-ACE scans a regular lattice of excitation spots, and employs individual pixels of a pixelated detector as a regular lattice of pinholes.
An image can be acquired by raster scanning the spots with width and height equal to the separation distance between focal points of the lattice, so that each spot scans an independent rectangular subregion of the total field of view.

Aberrations correction is performed independently for the excitation light and fluorescence emission. Excitation light is corrected anisoplanatically through computer generated holography, while fluorescence is corrected through conventional, isoplanatic adaptive optics by use of a DM.

In order to generate a lattice of independently generated corrections in different spots, an SLM, conjugated to the back aperture of the system, is used to modulate the phase of excitation light with a computer generated hologram (CGH). 
In particular, In order to generate $N$ spots, each located at positions $x'_n,y'_n$ and each corrected by a pupil aberration $\phi_{AO,n}$, the CGH can be computed as:
\begin{equation}
\phi_{CGH}=arg{\sum_{n=1}^{N}{e^{i(\phi_{POS,n}+\phi_{AO,n}+\theta_n)}}},
\end{equation}
where $\theta_n$ are constant phase terms, which can be optimized to improve the hologram quality with a variety of algorithms \cite{di2007computer}.  $\phi_{POS,n}$ is a displacement phase pattern
\begin{equation}\label{eq:displacement phase}
\phi_{POS,n}(x,y)=\frac{2 \pi}{\lambda f}(xx'_n+yy'_n),
\end{equation}
where $\lambda$ is the excitation wavelength, and $f$ is the equivalent focal of the system.

\begin{figure}[ht]
\centering
\includegraphics[width=\textwidth]{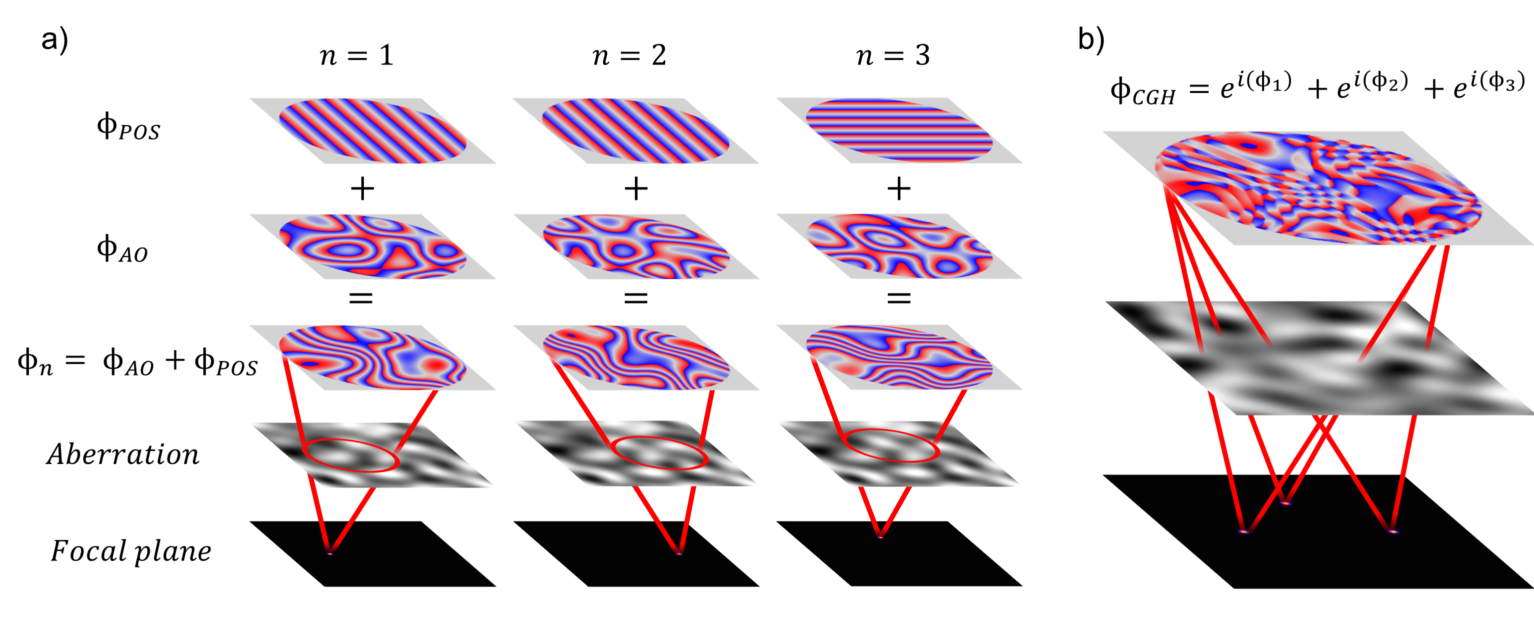}
\caption{\textbf{Independently corrected spots generation.} \textbf{a)} phase patterns generating single corrected spots in different locations in the sample. It should be apparent that the corrected phase aberration varies depending on the spot location. \textbf{b)} Multispot hologram with independent spots correction. The hologram is a complex sum of the phase patterns generating the single corrected spots. It is crucial to notice that all correction patterns are defined over the full size of the pupil, therefore exploiting the full resolution of the corrector.}
\label{fig: qualitativecorrection}

\end{figure}

The general principle of independently corrected CGH generation is qualitatively represented in Fig.~\ref{fig: qualitativecorrection}.

It is important to notice that, since the complete point spread function of a confocal fluorescence microscope is the product of the excitation light point spread function with the fluorescence emission point spread function \cite{visser1994electromagnetic}, correction of the excitation wavefront alone is sufficient for restoring optimal resolution.

Correction of the isoplanatic aberration of the fluorescence light, while not optimal for all points, can still improve the signal to noise ratio of the acquired image.

As a collateral, but useful addition, due to the independent correction of excitation and fluorescence light, LSFM-ACE provides a form of correction of chromatic aberrations between excitation and emission wavelengths.

\section{Sensorless optimization approach}
\label{sec:sensorless_opt}

While wavefront sensing based correction of excitation light is possible \cite{wilding2016adaptive}, its implementation in a parallelized focus method would be extremely complex, and exceed the scope of this paper. Instead, a simpler sensorless  approach was developed.

The sensorless correction of aberrations in LSFM-ACE is performed in two steps. Firstly, the DM is employed to perform conventional isoplanatic adaptive optics. In the specific case of the experimental setup used in this paper (described in section \ref{sec:setup}), the DM is located in the optical path shared by excitation and emission light, performing correction on both paths.
Secondly, independent correction of all excitation spots is performed through the SLM. 

Isoplanatic correction can be performed with any previously reported method. For the results presented in this paper, we use the optimal model based method previously developed by the authors \cite{pozzi2018optimal}. For this specific optimization procedure, the second moment of the average spatial distribution of the camera images of spots is used as a metric.

Anisoplanatic correction of a total of $N$ spots requires estimation of a single CGH $\phi_{CGH}$ simultaneously maximizing $N$ metric functions $\Upsilon_n(\phi)$, each measuring the correction performance in the area scanned by the $n$-th spot.

The metric functions $\Upsilon_n(\phi)$ were computed as the average intensities of the $N$ 1-dimensional fluorescence intensity signals acquired by using the DM to perform a circular scan of diameter equal to the distance between spots. A circular scan was preferred to a raster scan as it allows to compute metric values in a considerably shorter time.

For the presented results, the optimization implemented is a parallelized generalization of previously reported \cite{debarre2007image} methods, consisting on an hill-climb optimization over a gradient orthogonal Lukosz polynomial base.

\begin{figure}[ht]
\centering
\includegraphics[width=\textwidth]{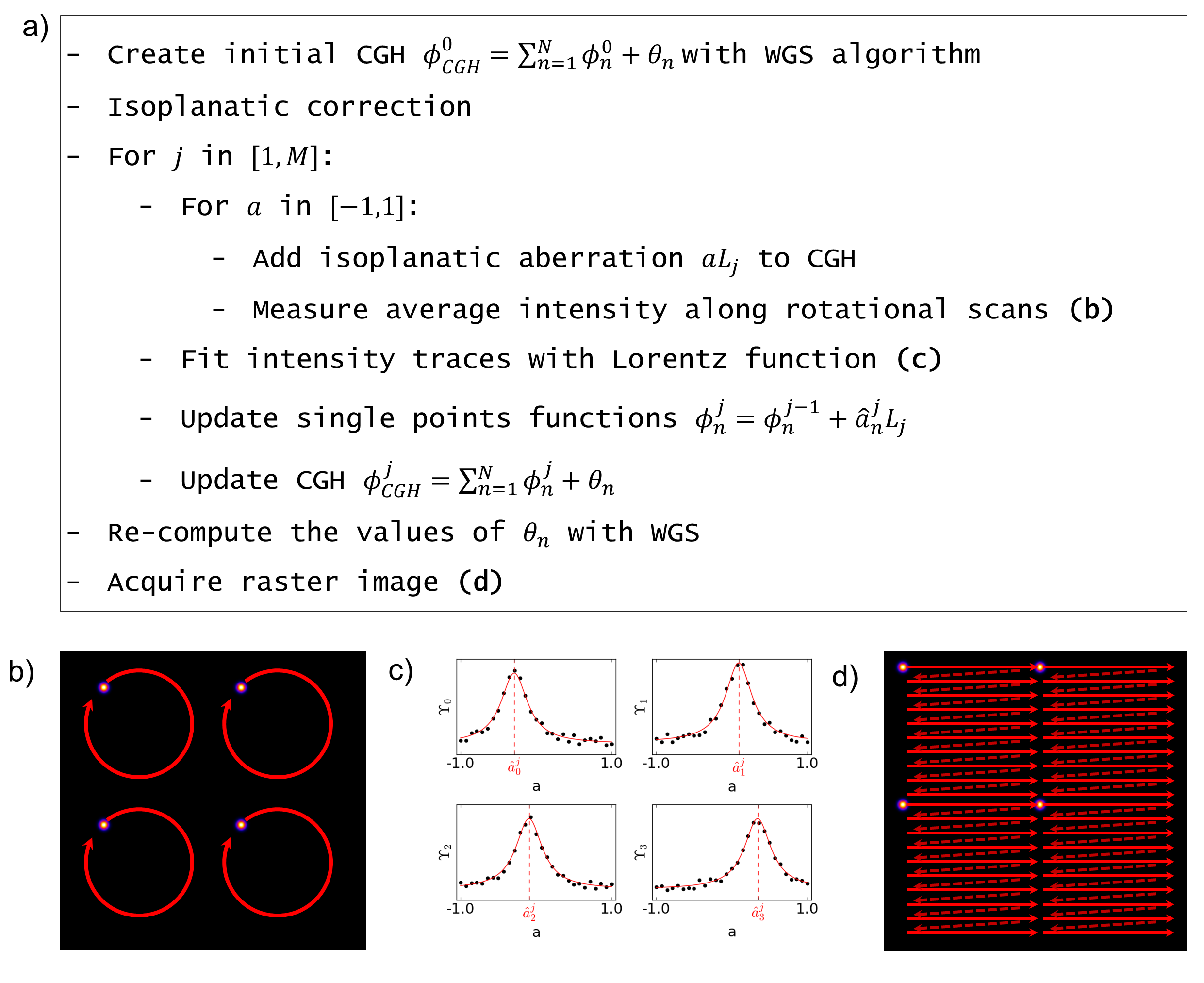}
\caption{\textbf{Brief overview of the microscope operation workflow.} \textbf{a)} algorithm for the acquisition of a anisoplanatically corrected image. \textbf{b)} schematic representation of the intensity distribution and scanning path at the sample plane for a 4 spot LSFM-ACE, with scanning path for the acquisition of metric values. \textbf{b)} Simulated examples of Lorentzian fits for the estimations of maximas of a given coefficient. \textbf{c)} Scanning path for the final image acquisition.}
\label{fig: algorithm}

\end{figure}

The optimization is initialized by computing a starting CGH $\phi_{CGH}^0$ generating a lattice of $N$ uncorrected spots:
\begin{equation}
\phi_{CGH}^0=arg{\sum_{n=1}^{N}e^{i(\phi_{n}^{0}+\theta_{n})}},
\end{equation}
where $\theta_{n}$ are constant phase terms, computed through the weighted Gerchberg Saxton algorithm \cite{di2007computer}, and $\phi_{n}^{0}$ are the phase patterns generating independently each of the $N$ spots, as reported in equation \ref{eq:displacement phase}. 

After initialization, the optimization is performed through a hill-climb procedure over a truncated base of Lukosz polynomials \cite{debarre2007image} $\{ L_1, ... ,L_M \}$.
At the $j$-th iteration of the algorithm, the metric functions $\Upsilon_n(\phi^{j-1}+aL_j)$ are measured for linearly modulated values of the amplitude $a$, in order to solve the $n$ one-dimensional maximization problems:
\begin{equation}
\max_{a^j_n}\Upsilon_n(\phi^{j-1}+{a^j_n}L_j) \ ,
\end{equation}
by fitting the functions $\Upsilon_n$ with a Lorentzian models \cite{debarre2007image}, in order to estimate the amplitudes $\hat{a}^j_n$ maximizing each function.
A new CGH is then computed as:
\begin{equation}
\phi_{CGH}^j=arg{\sum_{n=1}^{N}e^{i(\phi_{n}^{j}+\theta_{n})}} \ ,
\end{equation}
where:
\begin{equation}
\phi_{n}^{j}=\phi_{n}^{j-1}+\hat{a}^j_nL_j \ ,
\end{equation}
and the procedure is performed until $j=M$.

It is to be noticed that the values of $\theta_{n}$ are not re-optimized at each step of the procedure, in order to reduce the total correction time. This can reduce the uniformity of intensity between spots during optimization, but has no effects on the estimation of the values of the correction coefficients $\hat{a}^j_n$.
After the $M-th$ iteration, the values of $\theta_{n}$ are recalculated through the Weighted Gerchberg Saxton algorithm to ensure uniform intensity between spots, and the system is ready to acquire an aberration corrected image.

The described optimization approach was chosen due to its simplicity in implementation for simultaneous optimization of multiple functions. Better performances in optimization speed could, in principle, be achieved through the development of parallelized versions of other algorithms.

\section{Experimental setup}
\label{sec:setup}
\begin{figure}[ht]
\centering
\includegraphics[width=\textwidth]{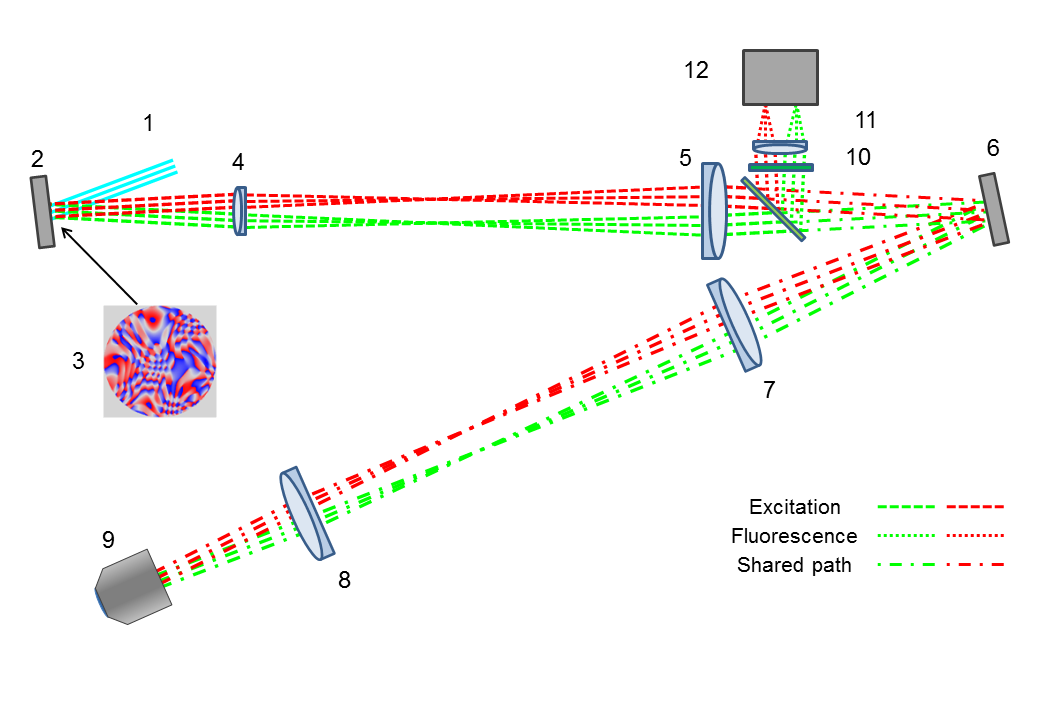}
\caption{\textbf{Optical setup.} Schematic representation of the optical setup, reporting the optical paths for two spots at the edges of the field of view, represented in different colors. Displacement between spots is exaggerated to aid readability.
Excitation light (1) is modulated by the SLM (2) with the computed hologram (3). The SLM plane is conjugated by a telescope (4,5) to the DM plane (6). A second telescope (7,8) conjugates the DM to the back aperture of the objective (9). Fluorescence light propagates the opposite direction, is descanned by the DM, filtered by a dichroic and emission filter set (10) and focused by a short focal tube lens(11) on the camera detector (12).}
\label{fig: Setup}
\end{figure}

A custom LSFM-ACE microscope was realized for green fluorophores imaging. A schematic representation of the system is reported in figure \ref{fig: Setup}. Excitation light was provided by a single mode, solid state laser at $488 nm$ (Sapphire 488, Coherent) expanded to a beam waist of 10mm in order to provide uniform illumination of the SLM surface. The SLM employed (P512-0532, Meadowlark optics) has resolution of $512 \times 512$ pixels on a surface area of $7 \times 7 mm$, with a refresh rate of $200 Hz$.

A $1.5\times$ telescope of two achromatic doublets of focal lengths $100$ and $150 mm$ respectively was used to conjugate the surface of the SLM to the surface of the DM.
A vertically mounted thin copper wire was used in the focal plane of the first lens of the telescope to mask the zeroth order of diffraction of the SLM. In this configuration, the wire does not block any of the generated spots, as long as they are organized in a lattice of even number along the horizontal axis.

A 69 actuators voice coil based DM (DM-69, Alpao) was used both for isoplanatic aberration correction and scanning, therefore preventing the need for an additional scanning device.

A $0.66\times$ telescope of two achromatic doublets of focal lengths $150$ and $100 mm$ respectively was used to conjugate the surface of the DM to the back aperture of the microscope objective ($40\times$, $1.0 N.A.$, water dipping, Zeiss).

A dichroic mirror ($496 nm$ shortpass, Semrock) was positioned in infinity space between the DM and the second lens of the telescope conjugating SLM and DM. Fluorescence light was further filtered by a bandpass filter (MF525-39, Thorlabs). A $75mm$ achromatic lens was used to focus images of the spots to an sCMOS camera (Optimos, QImaging).

In order to correct for aberrations due to the optical system itself, a baseline correction was performed by imaging a test sample of fixed bovine pulmonary artery endothelial cells with Bodipy staining on $\alpha$-tubulin (Fluocells slide $2$, Invitrogen). Due to the very thin and trasparent nature of the sample, such baseline correction was assumed to be the system aberration, and applied to all acquired images.

\section{Results}

The system was tested on fixed whole Drosophila Melanogaster brains, expressing green fluorescent protein on isolated neurons (see \ref{sec:methods}). A lattice of $8 \times 8$ spots was generated with the SLM, each scanning a $16 \times 16 \mu m$ area in a $100 \times 100$ pixels raster pattern, achieving sampling at the objective diffraction limited resolution. Pixel intensities were sampled by the camera at $400 Hz$. This resulted in images with a field of view of $128 \times 128$ micron, with a resolution of $800 \times 800$ pixels with an acquisition time of approximately $30s$.

Isoplanatic adaptive optics was performed on a set of 24 gradient orthogonal \cite{pozzi2018optimal} mirror modes. Correction time for a single image was approximately $5s$.

Anisoplanatic correction was performed as described in section \ref{sec:sensorless_opt}. The first 24 Lukosz polynomials, excluding tip,tilt and defocus, were used as modes for correction.
While the SLM is capable of correcting much higher order aberration, a set of 24 low order aberrations was chosen in order to ensure optimization times compatible with experimental needs. Moreover, this ensures that anisoplanatic and isoplanatic corrections are performed on aberrations of similar order, ensuring that any improvement in image quality is due to anisoplanatic correction, and not to correction of higher order aberrations.
The metric was acquired on circular scans consisting of 20 data points each. 29 circular scans were acquired for each mode. In experimental condition, anisoplanatic correction of a single plane required approximately 1 minute, including the final Weighted Gerchberg Saxton optimization step.

\begin{figure}[ht]
\centering
\includegraphics[width=\textwidth]{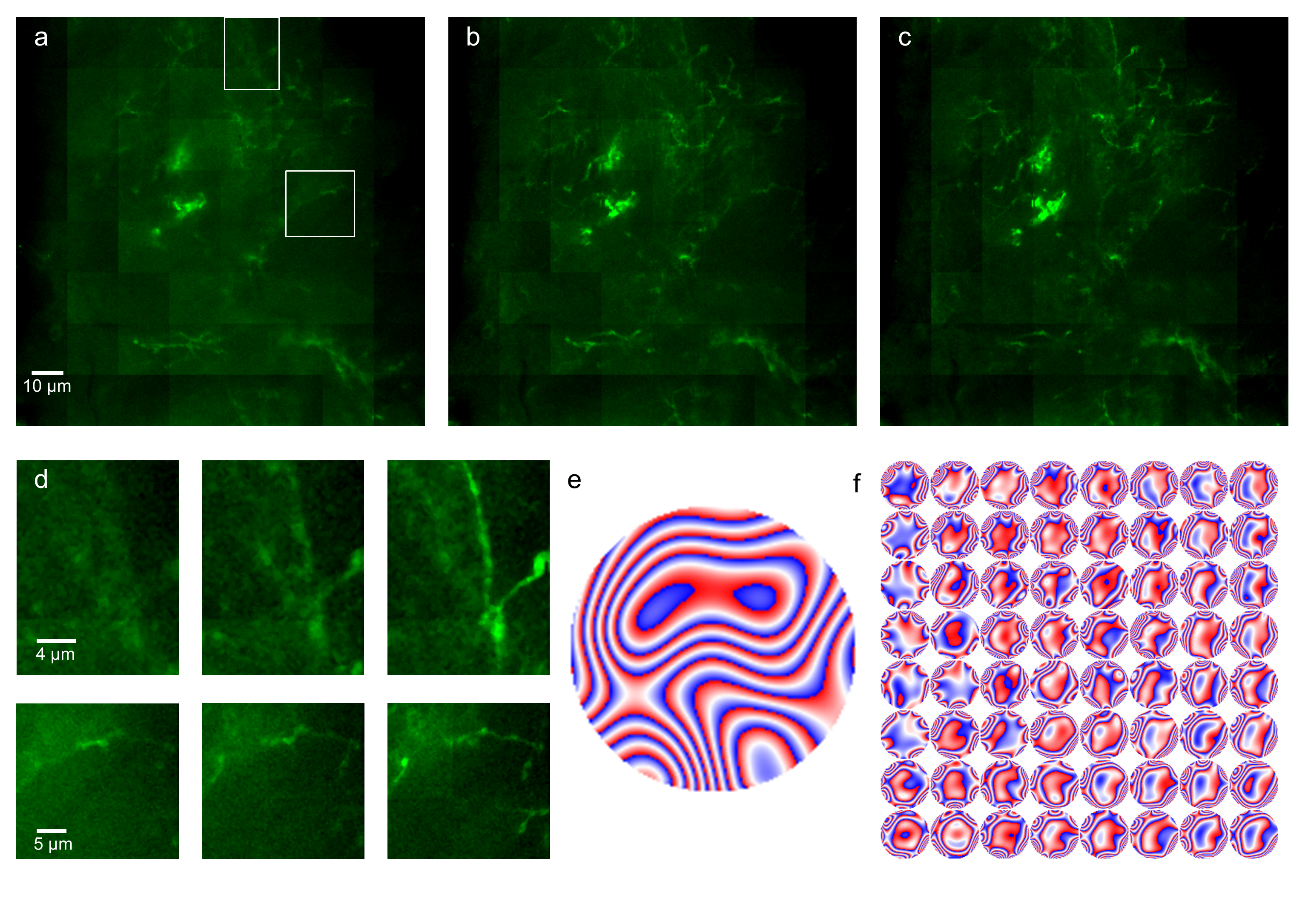}
\caption{\textbf{Single image correction comparison.}Example of image acquired with LSFM-ACE. \textbf{a) } Image without adaptive optics, \textbf{b) } Image corrected with isoplanatic adaptive optics, \textbf{c) } Image acquired with anisoplanatic correction of excitation. \textbf{d) } Detail from the areas highlighted with rectangles in panels a, b and c, with the three methods of correction. \textbf{e) } wrapped isoplanatic aberration corrected in b, \textbf{f) } wrapped corrections applied to individual isoplanatic patches in c.}
\label{fig: bigcomparison}

\end{figure}

In order to compare performances between the uncorrected system, the isoplanatic and the anisoplanatic correction, each image was acquired first without correction, then with isoplanatic correction, and finally after anisoplanatic optimization. This was done so that, in case of appearence of photobleaching, no unfair advantage in the comparison would be given to the presented method. Nonetheless, no significant photobleaching was observed during the acquisition of the images reported in this paper.

Figure \ref{fig: bigcomparison} reports an example image acquired without correction, with isoplanatic correction, and with LSFM-ACE. It can be notice that, while traditional isoplanatic correction greatly enhances the quality and sharpness of the image, the introduction of an anisoplanatic correction component on the excitation greatly improves contrast, and allows the resolution of details of the image which would simply be lost due to aberrations.

Panel \textbf{e} and \textbf{f} of figure \ref{fig: bigcomparison} report the final corrected isoplanatic  and anisoplanatic aberrations. It can be observed that the isoplanatic component has relatively low order, while the single corrections of individual patches in the LSFM-ACE image present high order modes and widely vary across the field of view.

\begin{figure}[ht]
\centering
\includegraphics[width=\textwidth]{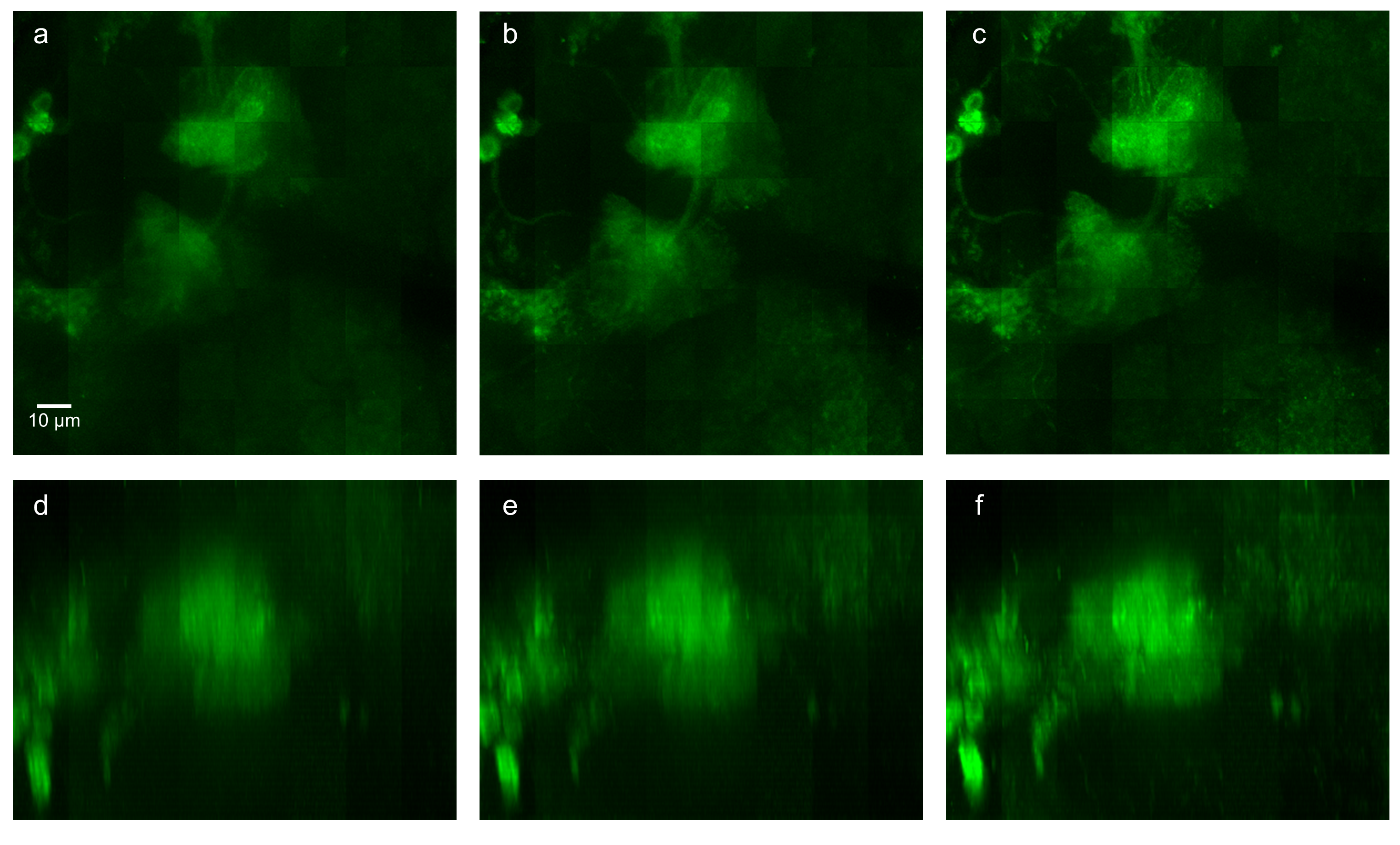}
\caption{\textbf{Maximum intensity projections of 3d images.}Samples were scanned over their full depth, computing new corrections every $15 \mu m$ of depth. \textbf{a) }XY image without correction \textbf{b)}XY image with isoplanatic adaptive optics \textbf{c)}XY image with anisoplanatic correction. \textbf{d,e,f)} XZ projections the same datasets. }
\label{fig: mips}
\end{figure}

In order to evaluate performance of the correction in three dimensional imaging, axial scans of the sample over $10 \mu m$ were acquired, updating the correction every $10 \mu m$. Figure \ref{fig: mips} report maximum intensity projections of the images, and some details. It can be observed how isoplanatic adaptive optics can not recover the full details in the image, which are only visible after anisoplanatic correction.
A three dimensional reconstruction of the full sample is available as supplementary material.

\section{Discussion}

In this paper, a new method for the correction of sample induced aberrations in confocal microscopy has been presented.
The presented method exploits computer generated holography to illuminate the sample with a lattice of independently corrected excitation spots, which can be raster scanned through the sample to generate a parallelized confocal microscopy image. A deformable mirror is used to perform conventional isoplanatic adaptive optics on the fluorescence light.

The results presented show how the isoplanatic nature of conventional correction approaches severely limits the performance of adaptive imaging systems in biological samples, and the introduction of anisoplanatic correction greatly enhances the image quality.
If compared to existing anisoplanatic correction methods (e.g. conjugate adaptive optics \cite{mertz2015field,park2015high}, multiplexed pupil \cite{park2017large} or time multiplexing \cite{liu2018observing}), LSFM-ACE provides the first real parallelized correction without tradeoffs in correction resolution, isoplanatic patch size, or acquisition time. Moreover, by conjugating both phase modulators to the pupil plane of the objective, the optical design is simpler to design, implement, align and operate.

The method allows high resolution, optical sectioning imaging in relatively thick and turbid media potentially providing, in specific cases, an alternative to multiphoton microscopy when the intrinsic advantages of linear excitation are needed (e.g. higher resolution, lower cost, higher spectral separation of excitation.).

It is to be noticed how, while the method only corrects anisoplanatically the excitation light path, ensuring that a fully corrected excitation beam scans the sample greatly increases the image resolution. 
 It can be assumed that uncorrected anisoplanatic aberration in the fluorescence light path only produces a degradation in detected light intensity, but does not affect the image resolution.
 
 The signal limitation due to limited correction of fluorescence signal could be provided in the future by implementation of the method in multiphoton microscopy, provided that the isoplanatic correction applied to fluorescence light is sufficient for separation of the signals from different excitation spots on the camera.
 
 Finally, while the reported measurements were provided for low order aberrations, the presented method, being based on a high resolution corrector, could allow  higher order aberrations correction, up to the limit of scattering compensation \cite{vellekoop2007focusing}. This would, however, require faster refreshing SLM technologies, faster optimization algorithms \cite{verstraete2015model}, and some form of high order correction of fluorescence light.

\section{Methods}\label{sec:methods}
\textbf{Samples preparation.}We imaged Drosophila brains containing native GFP and mRFP signal. Frist, the brains were dissected in PBS (1.86 mM NaH2PO4, 8.41 mM Na2HPO4, and 175 mM NaCl) and fixed using a ice-cold 4\% paraformaldehyde solution for an additional 60 min at room temperature. The brains were washed 3x 10 minutes PBS solution with 0.1 \% Triton X-100 and again 2x 10 minutes washed with PBS. The native fluorescence signal was boosted by immunostaining against GFP and mRFP as previously described in \cite{perisse2016aversive}; primaries at 1:2,000 for 2 overnights at 4C and secondaries at 1:500 for 1 overnight at 4C. We used anti-GFP (chicken, abcam13970) and anti-DsRed (Rabbit, Clontech 632496, 1:2,000). 

\section*{Author contributions}
P.P. Ideated the technique, realized the experimental setup and performed the measurements. C.S. prepared and provided the samples. D.W. partially contributed to computer code for the experiments. O.S. ideated the isoplanatic correction method implemented.
P.P. wrote the manuscript, with proofreading and contribution by all other authors. Management and funding for the microscopy facility was provided by G.V. and M.V. , management and funding for the samples preparation facilities were provided by M.B.

\section*{Acknowledgments}
The research leading to these results has received funding from the European Research Council under the European Union's Seventh Framework Programme (FP7/2007-2013) / ERC grant agreement n. 339681. \\
 The work of G. Vdovin and O. Soloviev is partially funded by the program ``5 in 100'' of the Russian Ministry of Education, and by Flexible Optical B.V. . The authors would like to acknowledge the contributions of W.J.M. van Geest and C.J. Slinkman.

\section*{Competing interests}
The authors and funding institutions have no competing interests to declare.

\end{document}